\documentclass[manuscript]{aastex}

\shorttitle{LiH depletion}
\shortauthors{Bovino et al.}
\begin{document}
\title{Fast LiH destruction in reaction with H: quantum calculations and astrophysical consequences.}
\author{S. Bovino, M. Wernli and F. A. Gianturco}
\affil{Department of Chemistry and CNISM, the University of Rome "Sapienza", \\ P.le A. Moro 5, 00185 Rome, Italy}
\email{fa.gianturco@caspur.it}

\begin{abstract}
We present a quantum-mechanical study of the exothermic $^7$LiH reaction with H. Accurate reactive probabilities and rate coefficients are obtained by solving the Schr\"{o}dinger equation for the motion of the three nuclei on a single Born-Oppenheimer potential energy surface (PES) and using a coupled-channel hyperspherical coordinate method. Our new rates indeed confirm earlier, qualitative predictions and some previous theoretical calculations, as discussed in the main text. In the astrophysical domain we find that the depletion process largely dominates for redshift (z) between 400 and 100, a range significant  for early Universe models. This new result from first-principle calculations leads us to definitively surmise that LiH should be already destroyed when the survival processes become important. Because of this very rapid depletion reaction, the fractional abundance of LiH is found to be drastically reduced, so that it should be very difficult to manage to observe it as an imprinted species in the cosmic background radiation (CBR). The present findings appear to settle the question of LiH observability in the early Universe. We further report several state-to-state computed reaction rates in the same range of temperatures of interest for the present problem.
\end{abstract}

\section{Introduction}
The chemistry of the early Universe plays an important role in our understanding of the formation of the first chemical objects and on the birth and evolution of galaxies and interstellar clusters.  Observation of molecular absorption and emission lines have provided important information about the physical conditions in which molecules are first formed then excited and destroyed. Lepp et \emph{al} (2002) identified 200 reactions considered to contribute to the abundance of 23 atomic and molecular species. Molecular formation in the early Universe began when the temperature was low enough that the newly formed atoms could survive for further evolution. In this era, termed the recombination era, at a redshift around z = 2500 the Universe was about 100,000 years old and the temperature was about 4000 K. After recombination, the density was still very low and three-body (3B) reactions were still very inefficient: however, it was then that the first molecular species were postulated to be formed through radiative association. In spite of the low fractional abundance which is expected to exist for LiH ($\sim$10$^{-10}$-10$^{-19}$), this molecule has nevertheless been considered one of the most likely  species to be important in that domain, due to its large permanent dipole moment and to the light masses of the involved atoms. Its search has therefore spurred several experimental studies in the recent past (Combes and Wiklind, 1998; de Bernardis et \emph{al}, 1993). Indeed, it was further suggested by Dubrovich (1993) that primordial molecules with large dipole moments might be detectable by their imprint left on the CBR and arising from the resonant enhancement of the Thomson scattering cross sections which occur at the transition frequencies of the molecules. LiH was thus considered to be very relevant because of the role it may have played as a possible coolant during the late stages of  the gravitational collapse of the first cosmological objects, as this very molecule may have survived even H$_2$. In their works Lepp and Shull (1984), and later Puy et \emph{al} (1993) and Palla et \emph{al} (1995), concluded that the major formation mechanism of lithium hydride is likely to be via the following radiative association process:
\begin{equation}
\rm Li + H \rightarrow LiH + \nu
\end{equation}
The rate coefficients for reaction (1) were later calculated by Dalgarno et \emph{al} (1996) and Gianturco and Gori-Giorgi (1997), who considered also  the reduction of the abundance of LiH as a consequence of the exchange reaction, the latter being the strongly exothermic channel of the LiH destruction indicated below:
\begin{equation}
\rm LiH + H \rightarrow Li + H_2
\end{equation}
Since the observational detection of LiH is still an  ongoing possibility among the options of experimental efforts, the present work intends to clearly, and accurately assess the feasibility of such measures vis \'a vis the outcomes of rigorous computational data.
We therefore intend to study the very reaction (2) in the redshift range 1000 $<$ z $<$ 20 to try to provide a more detailed evaluation of the possible role of this molecule in the early Universe evolution. Quantum coupled-channel equations have been solved to first calculate accurate rate coefficients as described in section 2, while the astrophysical consequences of our results will be detailed in sections 3 and 4, together with our conclusions.
\section{The quantum reactive calculations}
In the last 20 years a great deal of effort has been spent to calculate accurate rate coefficients for the reaction (2). Stancil et \emph{al} (1996) conjectured a temperature-independent rate coefficient of 2 $\times$ 10$^{-11}$ cm$^{3}$s$^{-1}$ by assuming the absence of an activation barrier and therefore postulated the insignificant role of LiH in erasing primary anisotropies in the CBR field. Dunne et \emph{al} (2001), using a quasi-classical dynamics and Petrongolo et \emph{al} (2005), with a real wave-packet  method, obtained rate coefficients in a range of temperatures 0 $<$ T $<$ 4000 K. However, the potential energy surface (PES) used by those calculations (the one generated by Dunne et \emph{al}, 2001) presented a rather deep well in the product valley which would cause unphysical trapping and resonance effects in the dynamics within partners which are instead interacting via mainly van der Waals forces. We therefore carried out new calculations solving the quantum coupled-channel hyperspherical equations involving a new 3D-surface (Wernli et \emph{al}, 2008) recently calculated and fitted with very good accuracy. The exothermicity of the reaction was estimated to be 2.258 eV, and it now correctly proceeds without a barrier into the exothermic products of eq. (2). 

The main improvements introduced by our newly calculated PES for the ground electronic state of reaction (2) have been the correct description of long-range forces among product partners and the elimination of barrier features in the entrance channel of the reagents, as discussed in more detail by Wernli et \emph{al} (2008).

The reaction probability and rate coefficients were thus calculated for J = 0 using the reactive program ABC (Skouteris et \emph{al}, 2000) and carefully testing all convergence-controlling computational parameters in order to yield a confidence level on the final cross sections below 1\%. Following earlier suggestions (Zhang and Zhang, 1999) we further applied a uniform J-shifting approach to obtain the full rate coefficients from our knowledge of the J=0 rate ($\alpha^{J = 0}(T)$):
\begin{equation}
\alpha(T) \simeq \frac{k_B T}{B} \alpha^{J = 0}(T)
\end{equation}
where T is the temperature, $k_B$ is the Boltzmann constant and B is obtained by evaluating the rotational constant of the collision complex at the estimated geometry of the transition state of our PES. In our case B = 0.84 cm$^{-1}$, a value which slightly differs from the earlier suggestions (Padmanaban et \emph{al}, 2004). The main physical reasons for choosing the above approximation to generate the full rate coefficients rest on the reasonable assumption that higer $J$ values are chiefly modifying the shape of the transition states through changes of the centrifugal potential of the reaction complex. This is certainly a realistic description of an exothermic reaction that proceeds without a barrier in the entrance channels, as is indeed the case here.

\section{Results and discussion}
To start assessing the outcomes of the quantum calculations described in the previous section, we report in table 1 the LiH reaction rates, summed over all final states, for the formation of both oH$_2$ and pH$_2$ as product molecules. The range of temperatures covers all the values expected to be relevant for the corresponding range of z of astrophysical interest. Several comments could be made about these results:
\begin{itemize}
\item
the rates turn out to be rather large over the whole range of T and to be largely independent of yielding either pH$_2$ or oH$_2$ as products. Most importantly,  our new values from accurate ab initio methods turn out to be close to the educated estimates of Stancil et \emph{al} (1996), although extending and improving on their temperature dependence. We therefore  confirm with the present calculations the likely occurrence of a rapid disappearance of any newly formed LiH because of its reaction with the surrounding hydrogen. This is an important result which now comes exclusively from non-empirical calculations;
\item
above 200 K the rates were obtained by using a functional form $\alpha$(T)=aT$\times$exp(-bT), as suggested by Stancil et \emph{al} (1996) in their  kinetic model of lithium reaction in the primordial gas. Our calculated parameters are a=2.05$\times$10$^{-12}$ cm$^3$ s$^{-1}$K$^{-1}$ and b=0.00058 K for oH$_2$ and a=2.18$\times$10$^{-12}$cm$^3$ s$^{-1}$K$^{-1}$ and b=0.00084 K for pH$_2$. The values employed by Petrongolo et \emph{al} (2005) and by Padmanabam and Mahapatra (2002) were obtained by using the same formula but turned out to be slightly different and to yield different final rates. However, given the fact that they employed a different reaction PES (from Dunne et \emph{al}, 2001) this result is not surprising.
\end{itemize}
The further calculations reported by table 2 show another interesting facet of the present reaction since we give there the low-T rates obtained for the formation of oH$_2$ starting from different vibrational levels of the LiH molecule; one clearly sees that the initial process of LiH destruction changes when the LiH reacting molecule is internally "hot". Such states could be the result of the occurence of only partial relaxation of the latter species whenever is formed during the recombination process indicated by eq. (1). In any event, the order of magnitude of the rates does not vary dramatically when we change the vibrational energy content of the initial molecular partner. The results of the tables are pictorially given by the two panels of figure 1, where we show on the top panel the temperature dependence of the LiH destruction rates leading to the formation of both pH$_2$ and oH$_2$ and, in the lower panel, the oH$_2$ formation (caused after the LiH destruction) for different vibrational content of the initially formed LiH partner obtained through eq. 1. One sees there that the oH$_2$ formation is the favoured process outside a very small T-values region, as one should expect to find by nuclear multiplicity consideration. Furthermore, the presence of vibrationally "hot" LiH molecules can cause the depletion rates to become smaller, in favour of the simpler hydrogen exchange reaction.

The data of Table 3 additionally present the computed rates for the depletion reaction (2) leading to the formation of H$_2$ molecules in different, final rotovibrational states. This information is important for establishing the differences in the probabilities of producing internally excited hydrogen molecules which in turn can decay by radiative processes. The results cover the range of temperatures that also map the redshift values of relevance for the early Universe study, and clearly suggest that the individual H$_2$ formation rates change only little by changing the final rotovibrational level of the newly formed molecule: the examples of Table 3, in fact, indicate a variation of about one order of magnitude from forming H$_2$ ($\nu'$=0,$j'$=1) to producing instead H$_2$ ($\nu'$=3, $j'$=9). One should also note that the H$_2$-formation reaction becomes endothermic for the situations where it is forming H$_2$ in vibrational states above $\nu'$=4 (see also Wernli at \emph{al}, 2008).

The general behaviour shown by the data of Table 3 is further presented in a pictorial form by the two panels of figure 2, where we show as a function of the temperature, and of the redshift values, the relative behaviour of a selection of rates of H$_2$-formation into different, final rotovibrational states for the H$_2$ molecule. All rates appear to behave very similarly and to largely remain of the same order of magnitude, although we see that the production of internally excited species, e.g. see the formation of H$_2$ ($\nu'$=0, $j'$=11) in the upper panel and of H$_2$ ($\nu'$=3, $j'$=9) in the lower panel, corresponds to smaller rates. Hence, the present, state-specific results indeed suggest that the newly formed H$_2$ molecules should be preferentially formed in their ground rotovibrational level, with a reduced probability for further contributing to radiative emissions.

The relative behaviour of the two chief processes, i.e. the hydrogen exchange reaction (termed here LiH survival) and the H$_2$ formation (called the LiH destruction) are reported in figure 3 over a log-log scale in the main plot, while the same processes are shown on the temperature and redshift scales in the inset. The energy dependence clearly indicates that the H exchange is the dominant process at low collision energies, as expected because of the steric hindrance created by the bulkier Li atom to the low-T insertion reaction, while the destruction of the LiH initially formed is seen to occur for the majority of the collisional events once the temperature (and the energy) becomes larger than about 300 K. In this temperature range, as we can see in the inset, z is estimated (Puy et \emph{al}, 1993) to be greater than 250. This means that the destruction process is dominant in the main molecular formation range; we can then assert that, since the fractional abundance is dramatically reduced, the exchange reaction is likely to be negligible. Furthemore, we can assert that the high temperature processes markedly favour the destruction of any LiH formed during the recombination stage and therefore that no LiH species are expected to significantly survive that stage.

The final figure 4 reports now the general behaviour of the quantum reaction rates for the destruction process as a function of redshift, a quantity which covers the expected range relevant for early Universe evolution. One clearly sees once more that the rates around z=500 are indeed fairly large and unequivocally suggest that one should expect rapid disappearance of any newly formed LiH molecules by reaction with the H atom present in the astronomical environment during the recombination era.
\section{Present conclusions}
As is well known by now, the first galaxies and stars were formed from an atomic gas of chiefly H and $^4$He with trace amounts of D, $^3$He and $^7$Li. Since there was a clear absence of heavier elements, the necessary radiative cooling kinetics below 8000 K must have been controlled by a small fraction of that gas which was molecular, in order to make more efficient the collapse of the primordial clouds. In the present work we have therefore revisited the chemistry of the $^7$LiH molecule, expected to be formed via radiative recombination via eq. (1), to now understand from ab-initio calculations how likely it would be for it to survive the possible reactions with the very abundant hydrogen gas existing under those conditions. We have carried out quantum calculations of the reactive processes that are allowed at the existing conditions of those clouds and which essentially involve the exothermic channels of reaction (2), i.e. either the hydrogen replacement process LiH$^a$ + H$^b$ $\rightarrow$ LiH$^b$ + H$^a$, that allows the LiH molecule to survive, or the H$_2$ formation processes (into oH$_2$ and pH$_2$) LiH + H $\rightarrow$ Li + o,pH$_2$, which describe the destruction of the initial molecular partner.

The calculations employ a newly obtained PES (Wernli et \emph{al}, 2009) and carry out a coupled-channel study of the quantum reaction, summing over all possible final states of H$_2$, and further analysing the possible changes due to having the initial LiH molecule formed in some of its excited vibrational levels. The final rates were obtained by employing the uniform J-shifting procedure (Zhang and Zhang, 1999) and were extended to higher temperatures following the scheme outlined in section 2. The rates for both events (i.e. H replacement and H$_2$ formation) were thus obtained over a very broad range of redshift values, covering those that pertain to the recombination era.

The present calculations are the first quantum results in regions of astrophysical interest which employ both the "exact" coupled-channel dynamics and the best, thus far, reactive potential energy surface obtained from ab-initio computations (Wernli et \emph{al}, 2009). They allow us to make the following points:
\begin{itemize}
\item
over the range of relevant z  the destruction reaction dominates over the LiH "survival" of the hydrogen replacement channel (see figs 3 and 4);
\item
to have the initial molecule still excited in any of its internal vibrational states does not modify the above result, although showing an increase of the simple H-replacement rates as $\nu$ increases (see fig. 1);
\item
the analysis of the rates of H$_2$ formation into different rotovibrational levels (fig. 2) indicates that such species is preferentially formed into its lower ($\nu'$, $j'$) states, although the formation of internally "hot" H$_2$ molecules is far from negligible;
\item
formation of either oH$_2$ or pH$_2$ changes very little the computed rate values, thereby not influencing the above findings;
\item
the computed rates with the present PES turn out to be larger than the earlier calculations with a less accurate form of reactive interaction (Padmanaban, 2004; Petrongolo, 2005) and similar in size to the earlier, empirical estimates by Stancil et \emph{al}, 1996. Such accurate findings therefore indicate as being very hard to have and to observe any possible LiH survival under early Universe conditions.
\end{itemize}
In conclusion, the present new data strongly indicate rapid destruction of LiH molecules by exothermic reaction with the ubiquitous hydrogen atom in the clouds, a process very likely to occur at the time of the recombination era. As a consequence, it would be very difficult for such short-lived species to leave their observational imprinting in the CBR. One should also keep in mind that the present, accurate new depletion rates are directly affecting the use of cosmological models for estimating LiH abundances. As a consequence of such new values, the ensuing optical depth of the Universe due to elastic Thomson scattering of CBR photons from the present LiH species should be lower, and thus certainly different from that estimated in earlier studies (e.g. E. Bougleux and D. Galli, 1997): hence settling in the negative the question of possible LiH detection.

\acknowledgments
{\bf Acknowledgments}

We thank Alex Dalgarno for having suggested to us this problem long ago, for patiently waiting for our results and for carefully reading the present manuscript. We also thank E. Bodo for computing the new PES and for early discussions on this subject. The financial support of the University of Rome Research Committee, of the CASPUR Supercomputing Consortium and of the PRIN 2006 national project are gratefully acknowledged.

\begin{deluxetable}{ccc}
\tabletypesize{\scriptsize}
\tablecaption{Computed rate coefficients summed over all final states of the products. The notation 1.57 - 11 corresponds to 1.57$\times$10$^{-11}$.}
\tablewidth{0pt}
\tablehead{
\colhead{T(K)} & \colhead{$\alpha$(T)$_{ortho}$(cm$^3$s$^{-1}$)} & \colhead{$\alpha$(T)$_{para}$(cm$^3$s$^{-1}$)}}
\startdata
10 & 1.57 - 11 & 1.71 - 11\\
20 & 4.05 - 11 & 4.40 - 11\\
40 & 8.14 - 11 & 8.74 - 11\\
80 & 1.56 - 10 & 1.61 - 10\\
\ 100 & 1.94 - 10 & 2.00 - 10\\
\ 200 & 3.65 - 10 & 3.69 - 10\\
\ 500 & 7.66 - 10 & 7.16 - 10\\
\ \ 1000 &1.14 - 09  & 9.41 - 10\\
\ \ 2000 &1.28 - 09  & 8.12 - 10\\
\ \ 4000 &8.05 - 10  & 3.03 - 10\\ 
\ \ 5000 &5.64 - 10  & 1.63 - 10\\
\ \ 7000 &2.47 - 10  & 4.26 - 11\\
\ \ 8000 &1.58 - 10  & 2.10 - 11\\
\ \ \ 10000 &6.20 - 11  & 4.90 - 12\\
\enddata
\end{deluxetable}
\begin{deluxetable}{cccc}
\tabletypesize{\scriptsize}
\tablecaption{Calculated rate coefficients, summed over all final o-H$_2$ states, for different vibrational initial states of LiH.}
\tablewidth{0pt}
\tablehead{
\colhead{T(K)} & \colhead{$\alpha$(T)$_{\nu=0}$(cm$^3$s$^{-1}$)} & \colhead{$\alpha$(T)$_{\nu=2}$(cm$^3$s$^{-1}$)} & \colhead{$\alpha$(T)$_{\nu=4}$(cm$^3$s$^{-1}$)}
}
\startdata
10 & 1.57 - 11 & 3.33 - 13 & 6.83 - 13\\
20 & 4.05 - 11 & 1.48 - 12 & 2.27 - 12\\
40 & 8.14 - 11 & 6.28 - 12 & 4.29 - 12 \\
50 & 1.00 - 10 & 9.88 - 12 & 9.35 - 12\\
80 & 1.56 - 10 & 2.47 - 11 & 2.00 - 11\\
\ 100 & 1.94 - 10 & 3.72 - 11 & 2.90 - 11\\
\enddata
\end{deluxetable}
\begin{deluxetable}{cccccccc}
\tabletypesize{\scriptsize}
\tablecolumns{8}
\tablewidth{0pt}
\tablecaption{Computed rates of H$_2$ formation into different rotovibrational states. Only the oH$_2$ case is being presented. All units are cm$^{3}$s$^{-1}$.}
\tablehead{
\colhead{}    &  \multicolumn{3}{c}{$\nu'$=0} &   \colhead{}   &
\multicolumn{3}{c}{$\nu'$=3} \\
\cline{2-4} \cline{6-8} \\
\colhead{T(K)} & \colhead{$j'$=1}   & \colhead{$j'$=3}    & \colhead{$j'$=11} &
\colhead{}    & \colhead{$j'$=1}   & \colhead{$j'$=3}    & \colhead{$j'$=9}}
\startdata
\ \ 10 & 1.13-12 & 6.65-13 & 3.18-13 && 3.85-13 & 1.51-12 & 1.34-12 \\
\ \ 20 & 2.73-12 & 1.55-12 & 9.73-13 && 1.10-12 & 2.70-12 & 2.65-12 \\
\ \ 30 & 3.78-12 & 2.15-12 & 1.46-12 && 1.80-12 & 3.44-12 & 2.20-12 \\
\ \ 50 & 5.31-12 & 3.03-12 & 2.01-12 && 3.01-12 & 5.70-12 & 3.62-12\\
\ \ 70 & 6.68-12 & 3.86-12 & 2.30-12 && 4.15-12 & 9.04-12 & 3.85-12\\
\ 100 & 8.70-12 & 5.21-12 & 2.59-12 && 5.93-12 & 1.45-11 & 4.24-12\\
\ 120 & 1.05-11 & 6.15-12 & 2.79-12 && 7.15-12 & 1.80-11 & 4.52-12 \\
\ 200 & 1.29-11 & 7.86-12 & 2.48-12 && 1.21-11 & 3.78-11 & 2.73-12\\
\ 500 & 1.19-11 & 8.33-12 & 6.33-13 && 3.19-11 & 9.45-11 & 2.10-13\\
1000 & 4.51-12 & 3.91-12 & 2.81-14 && 6.97-11 & 1.92-09 & 1.27-13\\
\enddata
\end{deluxetable}
\begin{center}
\begin{figure}
\includegraphics[width=0.8\textwidth]{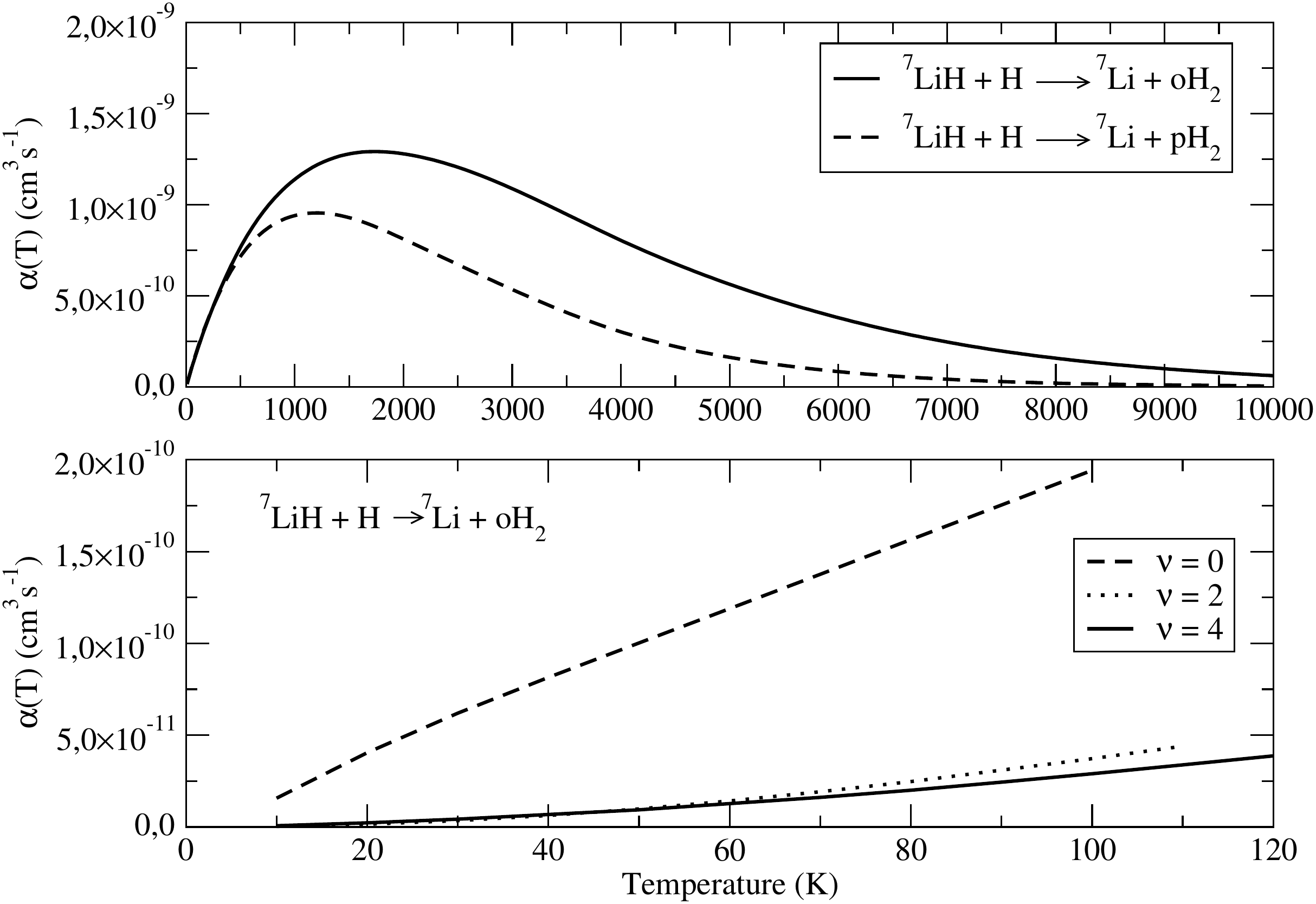}
\caption{Computed quantum reaction rates of LiH destruction. The top panel shows the dependence on the spin state of the final H$_2$ molecule over a very broad range of T, while the lower panel indicates, at the lower temperatures, the rate dependence on the initial vibrational state of LiH.}
\end{figure}
\end{center}
\begin{center}
\begin{figure}
\includegraphics[width=0.8\textwidth]{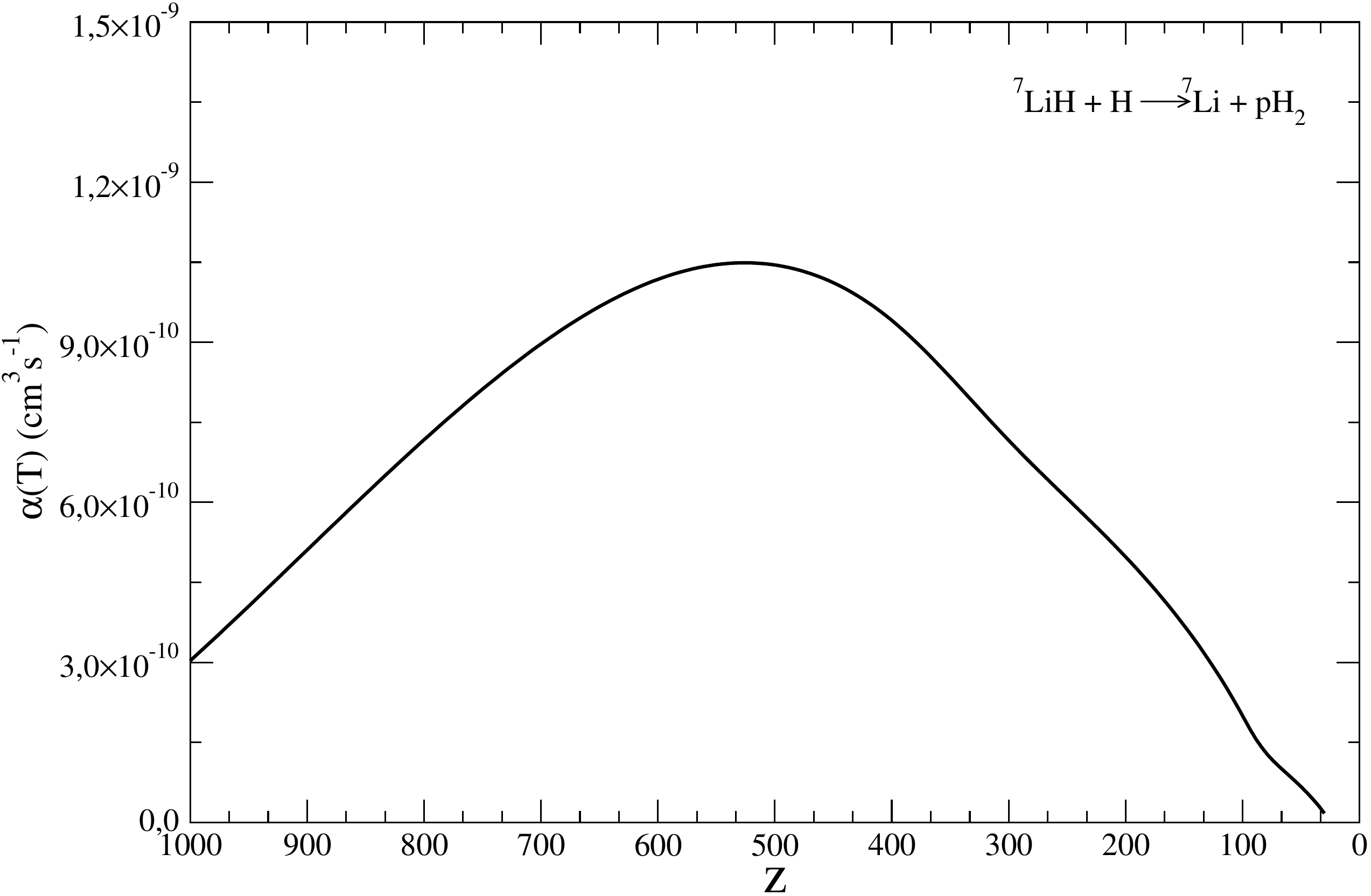}
\caption{Computed rates for H$_2$-formation reaction as a function of the temperature and of the redshift values, for different final, rotovibrational states of the H$_2$ molecule.}
\end{figure}
\end{center}
\begin{center}
\begin{figure}
\includegraphics[width=0.8\textwidth]{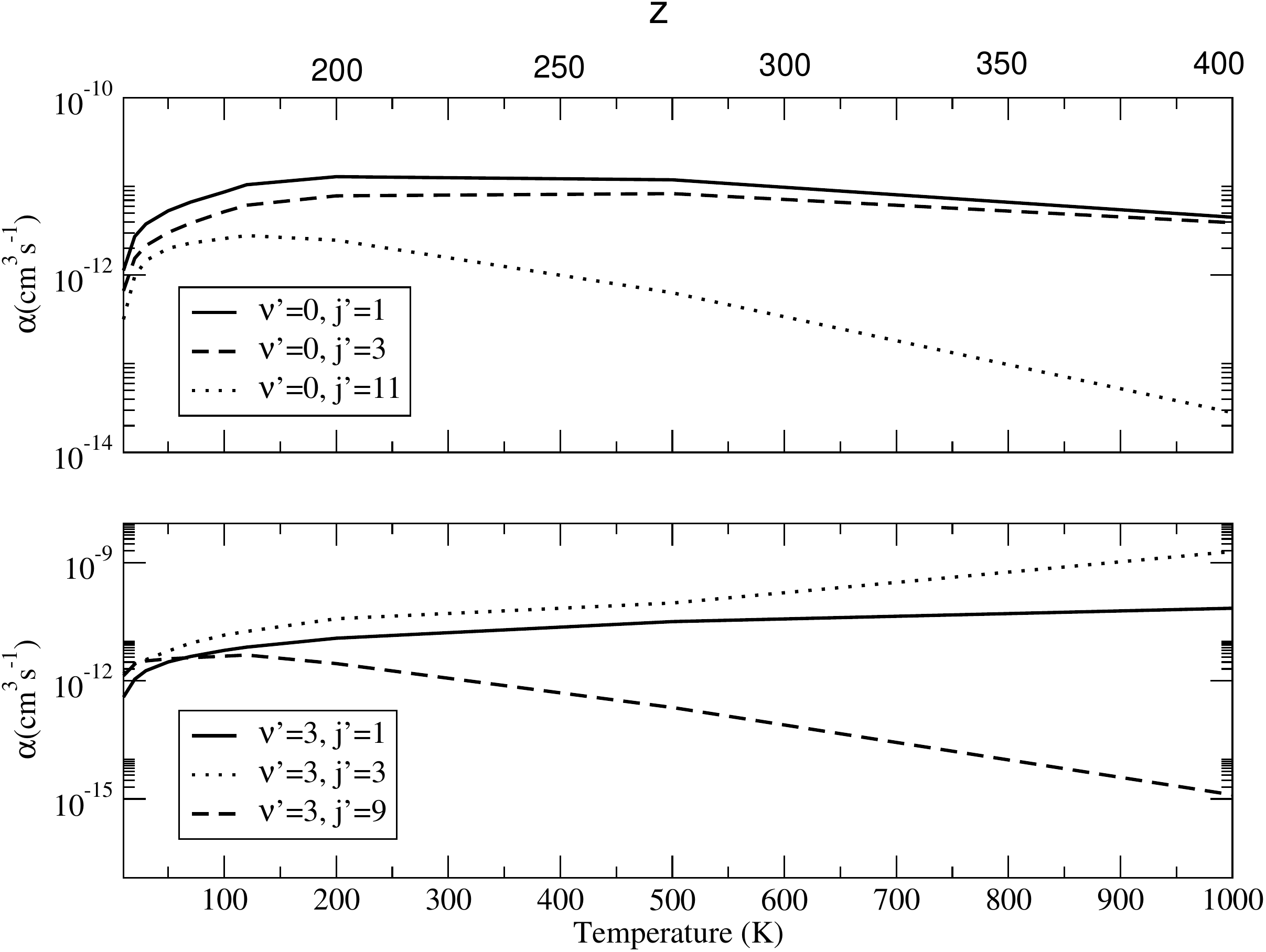}
\caption{Computed hydrogen-exchange reaction (dashed line) and o-H$_2$ formation (solid line) over a large energy range (in the main plot) and over the first 1000 K (in the inset).}
\end{figure}
\end{center}
\begin{center}
\begin{figure}[h]
\includegraphics[width=0.8\textwidth]{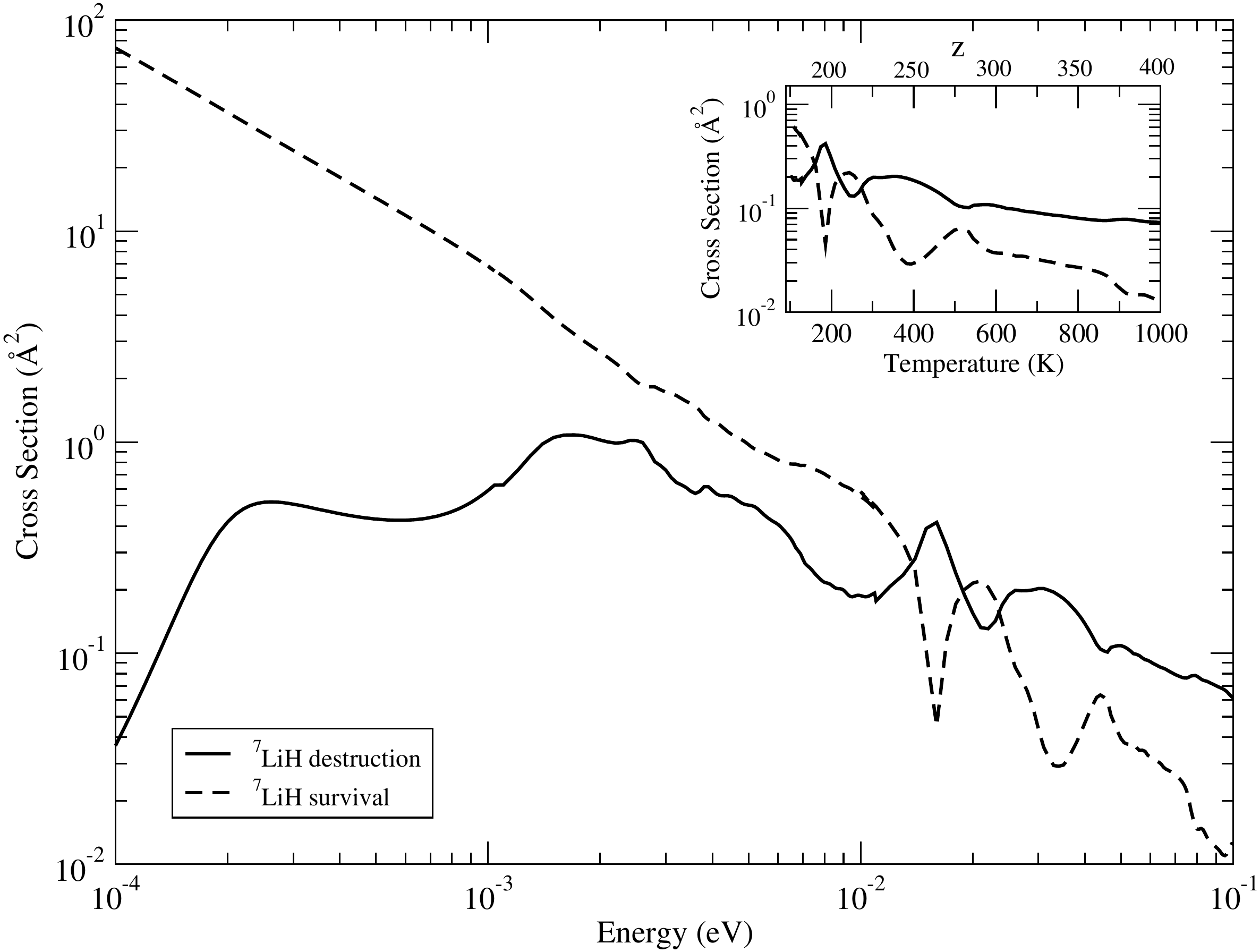}
\caption{Computed quantum LiH destruction rates, into p-H$_2$ formation, as a function of the z values.}
\end{figure}
\end{center}
\end{document}